\documentclass[preprint,tighten,aps]{revtex4}

\begin{document}

\baselineskip=1.1\baselineskip
\oddsidemargin=10mm
\parindent=1cm

\title
{\Large {Field theory and anisotropy of cubic ferromagnet near Curie point}}

\author{A. Kudlis$^{1,2}$,  A. I. Sokolov$^{1}$}

\email{ais2002@mail.ru}
\address
{$^{1}$Saint Petersburg State University, Saint Petersburg, Staryi Petergof,
Russia, \\ $^{2}$National Research University ITMO, Saint Petersburg, Russia}
\date{\today}

\begin{abstract}

{Critical fluctuations are known to change the effective anisotropy of
cubic ferromagnet near the Curie point. If the crystal undergoes phase
transition into orthorhombic phase and the initial anisotropy is not too
strong, effective anisotropy acquires at $T_c$ the universal value $A^* =
v^*/u^*$ where $u^*$ and $v^*$ are coordinates of the cubic fixed point
entering the scaling equation of state and expressions for nonlinear
susceptibilities. In the paper, the numerical value of the anisotropy
parameter $A$ at the critical point is estimated using the
$\epsilon$-expansion and pseudo-$\epsilon$ expansion techniques.
Pade resummation of six-loop pseudo-$\epsilon$ expansions for $u^*$,
$v^*$ and $A^*$ leads to the estimate $A^* = 0.13$ close to that
extracted from the five-loop $\epsilon$-expansion but differing
considerably from the value $A^* = 0.089$ given by the analysis of
six-loop expansions of the $\beta$ functions themselves. This
discrepancy is discussed and its roots are cleared up.}

\vspace{1.0cm}


\end{abstract}

\maketitle

\vspace{0.3cm}

\textwidth=15.5cm

Approaching the critical point thermal fluctuations of magnetization become so
strong that lead to the appreciable temperature dependence of effective
anisotropy of the cubic ferromagnet. To be precise, the crystalline anisotropy
of the nonlinear susceptibilities of different orders is meant here. If the crystal
undergoes a phase transition to the orthorhombic phase and its initial anisotropy is
not too strong, under $T \to T_c$ the magnetic subsystem either becomes isotropic
\cite{WF} or its anisotropy $A$ goes to a universal value which does not depend on
magnitude of $A$ far from the Curie point. For large initial values of $A$
fluctuations make the critical behavior of the system unstable and the first-order
phase transition occurs \cite{W,KW,LP}. Which one of the above-described modes is
realized in the critical region depends on the number of the order parameter
components (dimensionality) $n$. For $n < n_c$ the system goes to the isotropic
critical asymptote, while under $n > n_c$ it remains anisotropic at the Curie
point \cite{A}.

It is clear that the numerical value of the boundary dimensionality $n_c$
is of prime physical interest since it determines how real cubic ($n = 3$)
and tetragonal ($n = 2$) ferromagnets will behave in the vicinity of
$T_c$. The renormalization group (RG) analysis in the lower-orders of the
perturbation theory \cite {KW,S77,NR} as well as some lattice \cite{FVC}
and non-perturbative renormalization group \cite{TMVD} calculations have
shown that the $n_c$ likely lies between 3
and 4. However, further studies including the resummation of the
three-loop RG expansions \cite{MSI,Sh89} and multiloop RG calculations for
3D \cite{MSS,PS,CPV} and $(4 - \epsilon)$-dimensional \cite{KS} models
changed this estimate to $n_c < 3$
\cite{MSI,Sh89,MSS,PS,CPV,KS,KT,KTS,SAS,FHY1,FHY2,V,PVr}. In particular,
detailed 3D RG analysis \cite{CPV}, processing in different ways of the
five-loop $\epsilon$-expansions \cite{SAS,CPV} and addressing the
pseudo-$\epsilon$ expansion technique led to $n_c = 2.89$, $n_c = 2.855$,
$n_c = 2.87$, and $n_c = 2.862$, respectively. It means that cubic
ferromagnets undergoing the second-order phase transitions should belong to
the special -- cubic -- universality class, i. e. to possess a nonzero
anisotropy at the Curie point and to have their own specific set of
critical exponents.

On the other hand, the value of $n_c$ is very close to the physical value
$n = 3$ and thus the cubic fixed point should be located near the
Heisenberg fixed point at the phase diagram of RG equations. As a result,
the critical exponents corresponding to these points should almost
coincide with each other. The idea about degree of their closeness can be
obtained by comparing, e. g., the most reliable values of the
susceptibility exponent $\gamma$ for Heisenberg fixed point and the cubic
one; they are equal to 1.3895(50) \cite{GZ98} and 1.390(6) \cite{CPV},
respectively. It is quite clear that measuring of critical exponents in
experiments it is impossible to determine what kind of the critical
behavior is realized in cubic ferromagnets.

Nevertheless, to determine by means of physical and computer experiments
in what way the system with cubic symmetry behaves near the Curie point is
still possible but in order to do this one should study the nonlinear
susceptibilities of different orders $\chi^{(2k)}$ \cite{PS01}. As the RG
analysis has shown, the anisotropy of the nonlinear susceptibility
$\chi^{(4)}$ in the critical region may be as large as 5$\%$ what allows
to definitely consider it as a measurable quantity \cite{PS01}. At the
same time, this result can not be perceived as quite reliable because it
was obtained by processing of the series for $\beta$ functions of 3D cubic
model with coefficients that are not small and the expansions themselves
are known to be divergent.

In such situation it is natural to address the alternative theoretical
schemes which would allow to find the anisotropy of susceptibility
$\chi^{(4)}$ under $T \to T_c$ and to confirm or to correct the estimate
mentioned above. The $\epsilon$-expansion should be considered as the most
popular among such schemes; it is widely used for evaluation of critical
exponents and of other universal parameters of critical behavior. The
second field-theoretical approach being rather perspective in the
indicated sense is the pseudo-$\epsilon$ expansion technique proposed many
years ago by B. Nickel(see reference [19] in the paper \cite{LGZ80}). This
technique has demonstrated its exceptional efficiency when used to
estimate the universal characteristics of not only the 3D
\cite{LGZ80,GZ98,FHY2,HID04,NS14,SN14,NS15,SN16} but also of
two-dimensional systems \cite{LGZ80,COPS04,CP05,S2005,S13,NS13} for which
the known RG expansions are shorter and more strongly divergent.
Application of the pseudo-$\epsilon$ expansion approach accelerates the
convergence of the iterations and smoothes the oscillations of numerical
estimates so markedly, that in many cases for getting reliable
quantitative results it turns sufficient to use simple Pade approximants
or even to directly sum up the corresponding series.

Below, we find the value of the anisotropy of cubic ferromagnets near
the Curie point using $\epsilon$-expansion and pseudo-$\epsilon$ expansion
methods, compare the numbers given by three versions of field-theoretical
RG machinery mentioned above, and discuss the results obtained.

So, in the critical region the expansion of the free energy of the cubic
model in powers of magnetization components $M_{\alpha}$ can be written
down in the form:
\begin{eqnarray}
F(M_{\alpha}, m) &=& F(0, m) +  {\frac{1}{2}} m^{2 - \eta} M_{\alpha}^2 +
m^{1 - 2\eta}(u_4 + v_4 \delta_{\alpha \beta}) M_{\alpha}^2 M_{\beta}^2 +
... ,
\label{eq:1}
\end{eqnarray}
where $m$ is an inverse correlation length, $\eta$ being the Fisher
exponent, and $u_4$ and $v_4$ are dimensionless renormalized coupling
constants taking under $T \to T_c$ the universal values. In particular,
the fourth-order nonlinear susceptibility which is of interest for us can
be expressed via these couplings:
\begin{equation}
\chi_{\alpha \beta \gamma \delta}^{(4)} =
{\frac{\partial^3 M_{\alpha}}{{\partial H_{\beta}}{\partial H_{\gamma}}
{\partial H_{\delta}}}} \Bigg\arrowvert_{H = 0}.
\label{eq:2} \\
\end{equation}
Of special interest are the values of the nonlinear susceptibility for two
highly symmetric directions corresponding to the orientation of the
external field along the cubic axis ($\chi_{c}^{(4)}$) and the space
diagonal ($\chi_{d}^{(4)}$) of the unit cell. For these directions the
difference between the values of the $\chi^{(4)}$ is maximal, i. e. the
cubic anisotropy is most pronounced. As is easily to show,
\begin{equation}
\chi_{c}^{(4)} = - 24 {\frac{\chi^2}{m^3}} (u_4 + v_4), \qquad \quad
\chi_{d}^{(4)} = - 24 {\frac{\chi^2}{m^3}} \Biggl(u_4 + {\frac{v_4}{3}}
\Biggr),
\label{eq:3} \\
\end{equation}
where $\chi$ is a linear susceptibility. A role of natural characteristic
of the nonlinear susceptibility anisotropy plays the ratio
\begin{equation}
\delta^{(4)} = {\frac{|\chi_{c}^{(4)} - \chi_{d}^{(4)}|}{\chi_{c}^{(4)}}}.
\label{eq:4}
\end{equation}
This ratio
\begin{equation}
\delta^{(4)} = {\frac{2 v_4}{3(u_4 + v_4)}}
\label{eq:5}
\end{equation}
and can be expressed in terms of the anisotropy parameter
\begin{equation}
A = {\frac{v_4}{u_4}}. \label{eq:6}
\end{equation}
Its value at the Curie point $A^*$ we will find in this paper.

The fluctuation Hamiltonian of the cubic model has the form:
\begin{equation}
H = {\frac{1}{2}} \int d^D x \Biggl[ m_0^2 \varphi_{\alpha}^2 +
(\nabla\varphi_{\alpha})^2 + {\frac{u_0}{12}} \varphi_{\alpha}^2
\varphi_{\beta}^2 + {\frac{v_0}{12}} \varphi_{\alpha}^4 \Biggr],
\label{eq:7}
\end{equation}
where $\varphi_{\alpha}$ is $n$-component field of the order parameter
fluctuations, a bare mass squared $m_o^2 \sim T - T_c^{(0)}$, and
$T_c^{(0)}$ -- the phase transition temperature in the mean-field
approximation. The asymptotic value of the anisotropy parameter in the
critical region $A^*$ is determined by the coordinates of the cubic fixed
point of RG equations $u_4^*$ and $v_4^*$. For $D = 4 - \epsilon$ they are
known today in the five-loop approximation \cite{KS}. Corresponding
$\epsilon$-expansions for the physically interesting case $n = 3$ are as
follows:
\begin{equation}
u^* = 1.22222222\epsilon + 0.49291267\epsilon^2 - 0.4899848\epsilon^3 +
0.5912287\epsilon^4 - 1.429021\epsilon^5,
\label{eq:8}
\end{equation}
\begin{equation}
v^* = -0.40740741\epsilon + 0.19783570\epsilon^2 + 0.2270881\epsilon^3 +
0.1855026\epsilon^4 - 0.6908416\epsilon^5,
\label{eq:9}
\end{equation}
where  $u^* = (11/{2\pi})u_4^*$, $v^* = (11/{2\pi})v_4^*$. As is seen,
expansions (8) and (9) have an irregular structure (non-alternating),
their coefficients are not small and rapidly begin to grow with the number
of the series term. No wonder that the use of different resummation
methods including those based on the Borel transformation does not lead in
this case to more or less satisfactory results. Starting from the series
for the coupling constants, it is possible, however, to obtain the
$\epsilon$-expansion for the anisotropy parameter itself
\begin{eqnarray}
A^* = -\frac{1}{3} + 0.29629630\epsilon - 0.0673269\epsilon^2 +
0.458956\epsilon^3 - 1.31038\epsilon^4,
\label{eq:10}
\end{eqnarray}
which, being also divergent, has an advantage over (8) and (9) because it
is alternating. Since the higher-order coefficients of (10) strongly grow
up, it is reasonable to apply the Borel transformation for the processing
of this series, with the subsequent analytic continuation of the Borel
transform with a help of Pade approximants [L/M]. The results of the
resummation of (10) using this technique are presented in Table 1.
Although the estimates obtained are considerably scattered, the most
likely value of the $A^*$, i. e. that given by the diagonal approximant
[2/2], is close to the interval where the results of the six-loop RG
analysis are grouped. Indeed, the resummation of the series for $\beta$
functions of the three-dimensional cubic model leads to $u_4^* = 0.755 \pm
0.010$, $v_4^* =0.067 \pm 0.014$ \cite{CPV} what, in its turn, yields $A^*
= 0.089 \pm 0.018$. On the other hand, the distinction of the obtained
numbers -- 0.089 and 0.124 -- is not so small to completely ignore it.
This distinction should have origins. The specific feature of the
$\epsilon$-expansion for $A^*$ may be referred to as one of them: under
the physical value of $\epsilon$ first two terms of the series (10)
almost cancel each other what effectively shortens this series and
drastically worsens its approximating properties. Since
$\epsilon$-expansions for the coupling constants also have computationally
unfavorable structure we conclude that the method of $\epsilon$-expansion
does not allow to estimate with sufficient accuracy the magnitude of the
anisotropy of cubic ferromagnets in the critical region.

In such a situation it is reasonable to address the other version of the
RG perturbation theory -- the technique of pseudo-$\epsilon$ expansion.
The key idea of this method is to introduce, starting from the
three-dimensional RG expansions, a formal small parameter $\tau$ having
inserted it into linear terms of the series for $\beta$ functions and to
calculate observables as power series in $\tau $. Their numerical values
may be then obtained putting $\tau = 1$. So, we start from the six-loop RG
expansions for the cubic model \cite {CPV}. Iterating the equations
\begin{equation}
\beta_u (u, v) = 0, \qquad \qquad \beta_v (u, v) = 0
\label{eq:11}
\end{equation}
by the way just described, for the coordinates of the cubic fixed point
under $n = 3$ we obtain
\begin{eqnarray}
u^* = 1.2222222\tau + 0.3323342\tau^2 - 0.122585\tau^3
\nonumber\\
- 0.065595\tau^4 - 0.061083\tau^5 + 0.01269\tau^6,
\label{eq:12}
\end{eqnarray}
\begin{eqnarray}
v^* = -0.4074074\tau + 0.1306486\tau^2 + 0.232337\tau^3
\nonumber\\
+ 0.128399\tau^4 + 0.050252\tau^5 + 0.02224\tau^6.
\label{eq:13}
\end{eqnarray}

These series have a structure much more attractive than that of the
$\epsilon$-expansions (8), (9). Their higher-order coefficients are small
and decrease in modulo with increasing the number of the series term.
It may be expected that even direct summation of (12) and (13) will give
acceptable results. Indeed, substituting into the expansion (12) $\tau = 1$,
we get $u^* = 1.318$, the number which is close to the most reliable --
six-loop -- RG estimate $u^* = 1.321$ \cite{CPV}. For the second coupling
constant an analogous procedure gives $v^* = 0.1565$. This value considerably
differs from its six-loop RG counterpart $v^* = 0.117$ \cite{CPV} yielding,
however, for the anisotropy parameter the value 0.119 differing only slightly
from that extracted from the $\epsilon$-expansion for $A^*$.

Let's try to refine the estimates provided by the pseudo-$\epsilon$
expansion approach with a help of Pade approximants. Pade triangles for the
expansions (12), (13) are shown in Tables 2 and 3. It is seen that, although
three higher-order approximants for $u^*$ including [3/2] have
dangerous poles, the iteration procedure in general steadily converges to the
value 1.322 that is very close to the six-loop RG estimate $u^* = 1.321$
\cite{CPV}. This value will be accepted as an ultimate. The structure of the
content of Table 3 is also regular in the sense that the higher-order Pade
approximants yield the estimates of $v^*$ differing from the asymptotic
value $v^* = 0.1817$ by no more than $ 0.03 \div 0.04 $. Having found the
coordinates of the cubic fixed point we get $A^* = 0.1375$. This number is
certainly of interest: it strongly differs from the six-loop RG estimate
$A^* = 0.117/1.321 = 0.0886$ and is appreciably greater than the value
$A^* = 0.119$ obtained by direct summation of pseudo-$\epsilon$ expansions
for $u^*$ and $v^*$.

Further we will find an alternative estimate of the anisotropy parameter in the
critical region. It can be done by constructing the pseudo-$\epsilon$ expansion
directly for the $A^*$. Combining (12) and (13), we obtain:
\begin{equation}
A^* = -\frac{1}{3} + 0.197531\tau + 0.102951\tau^2 + 0.078982\tau^3 +
0.023907\tau^4 + 0.03847\tau^5. \label{eq:14}
\end{equation}
The coefficients of this series apart from the last one decrease in
modulo, the last coefficient is small and, therefore, the expansion (14)
is suitable for getting numerical estimates. The results of its summation
by means of Pade approximants are presented in Table 4. As is seen, two of
the highest-order approximants -- [4/1] and [1/4]-- have dangerous poles
while the asymptotic value $A^* = 0.1304$ is close to the number 0.1375
obtained above. Taking an average of these values and considering their
difference as a natural estimate of the accuracy of the approximation
scheme employed, we can assume that $A^* = 0.134 \pm 0.004$. On the other
hand, pseudo-$\epsilon$ expansions for $v^*$ and $A^*$ have unfavorable
feature: their first terms are negative and biggest in modulo, so the
numerical results are obtained as small differences of big numbers. It
certainly reduces their accuracy. Therefore we accept more conservative
estimate as a final one:
\begin{equation}
A^* = 0.13 \pm 0.01,
\label{eq:15}
\end{equation}
which seems to us realistic.

So, the pseudo-$\epsilon$ expansion method leads to the value of the
anisotropy parameter which 1.5 times greater than its RG analogue $A^* =
0.089$. How can one explain such a significant difference of two
field-theoretical estimates obtained within the highest-order available --
six-loop -- approximation? One of the possible reasons has already been
mentioned: the numerical value of $A^*$ is much smaller than the
coefficients of the first terms of the series employed, and it is
calculated as a small difference of big numbers. This smallness, in its
turn, is related to the fact that the boundary dimensionality of the order
parameter $n_c$ is close to 3. If $n_c$ coincided with the physical value
of $n$, then the anisotropy parameter would be equal to zero at the
critical point. Since the difference $3 - n_c$ is numerically small ($0.1
\div 0.15$), the value of $A^*$ turns out to be small as well. However,
precisely for this difference various field-theoretical schemes provide
significantly different estimates. For example, the processing of the
six-loop 3D RG expansions for $\beta$ functions of the cubic model by
means of the "conform-Borel" technique leads to $3 - n_c = 0.11$
\cite{CPV}, while the values of this difference obtained by the
resummation of the pseudo-$\epsilon$ and $\epsilon$ expansions for $n_c$
equal 0.138 \cite{FHY2} and 0.145 \cite{SAS}, respectively. Since the
first of the above mentioned numbers differs from the others by tenths of
percents, it is not surprising that a difference of values $A^*$, obtained
within the same iteration schemes, turns out to be so significant.

In conclusion we note that if one accepts the value of $A^*$ found above,
the anisotropy of the nonlinear susceptibility at the Curie point
\begin{equation}
\delta^{(4)} = {\frac{2 A}{3(1 + A)}}
\label{eq:16}
\end{equation}
will grow up almost to $8\%$. This makes the arguments in favor of the
possibility of experimental detection of the anisotropic critical behavior
in cubic ferromagnets \cite{PS01} even more convincing.

The work was supported by the Saint Petersburg State University (project
11.38.636.2013) and by the Russian Foundation for Basic Research (grant
15-02-04687).

\newpage

\begin{table}[t]
\caption{The Pade-Borel triangle for the $\epsilon$-expansion (10) of the
universal value of the anisotropy parameter $A^*$. Approximant [1/2] has
"dangerous" pole, so the corresponding number in the table is missing.}
\label{tab1}
\renewcommand{\tabcolsep}{0.4cm}
\begin{tabular}{{c}|*{5}{c}}
$M \setminus L$ & 0  & 1 & 2 & 3 & 4  \\
\hline
0 & $-$0.3333  & $-$0.0370  & $-$0.1044  &   0.3546  & $-$0.9558  \\
1 & $-$0.2063  & $-$0.0882  & $-$0.0482  &   0.0314  &          \\
2 & $-$0.1737  &      -     &    0.1241  &           &          \\
3 & $-$0.1568  & $-$0.1110  &            &           &          \\
4 & $-$0.1489               &            &           &          \\
\end{tabular}
\end{table}

\begin{table}[t]
\caption{The Pade triangle for the pseudo-$\epsilon$ expansion (12) of the
coupling constant $u^*$. Approximants [3/1], [3/2] and [2/2] have poles
close to 1, therefore the corresponding estimates are not reliable; in the
table they are bracketed. The bottom line (RoC) indicates the rate and the
character of convergence of Pade estimates to the asymptotic value. Here
Pade estimate of $k$-th order is the number given by corresponding
diagonal approximant [L/L] or by a half of the sum of the values given by
approximants [L/L$-$1] and [L$-$1/L] when a diagonal approximant is
absent. The resulting value of $u^* = 1.3218$ is obtained by averaging
over three working approximants [4/1], [2/3], and [1/4].} \label{tab2}
\renewcommand{\tabcolsep}{0.4cm}
\begin{tabular}{{c}|*{6}{c}}
$M \setminus L$ & 0 & 1 & 2 & 3 & 4 & 5 \\
\hline
0 & 1.2222  & 1.5546  & 1.4320    & 1.3664    & 1.3053   & 1.3180 \\
1 & 1.6787  & 1.4650  & 1.2909    & (0.4782)  & 1.3158   & \\
2 & 1.3545  & 1.3832  & (0.9025)  & (1.2483)  & & \\
3 & 1.3868  & 1.3629  & 1.3197  & & & \\
4 & 1.3004  & 1.3300  & & & & \\
5 & 1.3472  & & & & & \\
\hline RoC  & 1.2222  & 1.6166  & 1.4650  & 1.3371  & (0.9025) & 1.3218
\end{tabular}
\end{table}

\begin{table}[t]
\caption{The Pade triangle for pseudo-$\epsilon$ expansion (13) of the
coupling constant $v^*$. Approximant [1/1] has a pole close to 1,
corresponding estimate is unreliable and it is bracketed in the table. The
convergence of Pade estimates to the asymptotic value is illustrated by
the bottom line (RoC). Here the Pade estimate of $k$-th order is the
number obtained in the same manner as in the case of the constants $u^*$
(see Table 2).} \label{tab3}
\renewcommand{\tabcolsep}{0.4cm}
\begin{tabular}{{c}|*{6}{c}}
$M \setminus L$ & 0 & 1 & 2 & 3 & 4 & 5 \\
\hline
0 & $-$0.4074  &  $-$0.2768   & $-$0.0444  & 0.0840  & 0.1342  & 0.1565 \\
1 & $-$0.3085  & ($-$0.5753)  & 0.2426     & 0.1665  & 0.1741  & \\
2 & $-$0.2043  &    0.0416    & 0.1507     & 0.1729  & & \\
3 & $-$0.1505  &    0.1905    & 0.1906     & & & \\
4 & $-$0.1149  &    0.1906    & & & & \\
5 & $-$0.0900  & & & & & \\
\hline RoC & $-$0.4074 & $-$0.2926 & ($-$0.5753) & 0.1416 & 0.1507 &
0.1817
\end{tabular}
\end{table}

\begin{table}[t]
\caption{The Pade triangle for the pseudo-$\epsilon$ expansion (14) of the
anisotropy parameter in the Curie point. Approximants [2/1], [1/2], [4/1]
and [1/4] have dangerous poles, corresponding estimates are not reliable
and therefore are given in brackets. The bottom line (RoC) shows the
character of convergence of Pade estimates to the asymptotic value. Here
the Pade estimate of $k$-th order is the same as in Tables 2 and 3.}
\label{tab4}
\renewcommand{\tabcolsep}{0.4cm}
\begin{tabular}{{c}|*{6}{c}}
$M \setminus L$ & 0 & 1 & 2 & 3 & 4 & 5 \\
\hline
0 & $-$0.33333 & $-$0.13580 & $-$0.03285 & 0.04613 &  0.07004  & 0.10851 \\
1 & $-$0.20930 &    0.07921 &  (0.30639) & 0.08042 & (0.00689) & \\
2 & $-$0.14798 &  (0.25821) &   0.13630  & 0.13044 & & \\
3 & $-$0.10880 &   0.06129  &   0.13035  & & & \\
4 & $-$0.08417 &  (0.34520) & & & & \\
5 & $-$0.06592 & & & & & \\
\hline RoC & $-$0.33333 & $-$0.17255 & 0.07921 & (0.28230) & 0.13630 &
0.13039
\end{tabular}
\end{table}

\end{document}